%% file: main.tex
\documentclass{article}

\usepackage[utf8]{inputenc}
\usepackage{hyperref}
\usepackage{graphicx}
\usepackage[style = numeric-comp]{biblatex}
\input{MathCommands}

\usepackage{cleveref}
\usepackage{xspace,xcolor}
\usepackage[normalem]{ulem}

\bibliography{refs}

\title{Equity and Privacy: More Than Just a Tradeoff }
\author{
David Pujol\\ 
Duke University\\
dpujol@cs.duke.edu
\and Ashwin Machanvajjala\\
Duke University \& Tumult Labs\\
ashwin@cs.duke.edu
}

\begin{document}

\maketitle
\input{Introduction}

\input{Inequitites}
\input{Resolution}

\printbibliography
\end{document}

%% file: MathCommands.tex
\usepackage{amsmath,amsfonts,amsthm,bm}
\usepackage{hyperref,xspace}
\usepackage{xcolor,colortbl}

\newcommand{\eat}[1]{}

\def\eqref#1{equation~\ref{#1}}

\def\1{\bm{1}}

\DeclareMathAlphabet{\mathsfit}{\encodingdefault}{\sfdefault}{m}{sl}
\SetMathAlphabet{\mathsfit}{bold}{\encodingdefault}{\sfdefault}{bx}{n}



%% file: Introduction.tex

\section{Introduction}
Organizations large and small collect information about individuals and groups, and then  want to share insights learned from these sensitive data in order to inform research and policy making.
For instance, hospitals release de-identified data to innovate on disease detection and mitigation.
Internet companies train and release ML models as a service for a variety of classification tasks.
Government agencies release statistical data products to enable research and policymaking. 
However, releasing data, even in ``anonymous" or aggregated form, are vulnerable to privacy attacks that can result in sensitive properties of individuals in the data being revealed. 

There are a number of highly publicized privacy attacks on the release of anonymous or de-identified data. Individuals have been re-identified from anonymous search logs released by AOL, movie ratings released by Netflix, and taxi trips released by taxi companies. Membership inference attacks on machine learning models allow attackers to query ML systems and recover sensitive properties of individuals whose data were used to train the model. For instance, models trained on personal emails can reveal the content of those emails. Even the release of statistical data products can result in privacy breaches. In 2019, the US Census Bureau showed that summary statistics that were released as part of the 2010 decennial census (which were released using vetted statistical disclosure limitation methodology) could be combined to \textit{reconstruct} the sensitive database of records. Close to 50\% of the US population was reconstructed exactly down to exact age, race, and census block!

In response to these new and emerging threats, a number of privacy protection mechanisms have been proposed and deployed, culminating in \textit{differential privacy}, which is currently considered the gold standard for quantifying the privacy leakage of an individual due to a data release. Any privacy protection method performs some form of data minimization or perturbation -- such as redacting or coarsening attributes, sampling, noise infusion and so on -- which in turn reduces the quality/\textit{utility} of the data. And the stronger the privacy requirement, the lower the utility of the data release. 

While the entire field of privacy preserving data analytics is focused on the privacy-utility tradeoff, recent work has shown that privacy preserving data publishing can introduce different levels of utility across different population groups. It is important to understand this new tradeoff between privacy and \textit{equity} as privacy technology is being deployed in situations where the data products will be used for research and policy making. Will marginal populations see disproportionately less utility from privacy technology? If there is an inequity how can we address it? 

In this article, we examine this new tradeoff between privacy and equity. We begin by giving a brief history of modern privacy preserving techniques. We then discuss the privacy vs equity tradeoff using case studies from the recent literature on privacy preserving allocation and privacy preserving machine learning. Finally we discuss 
methods that may limit the inequity introduced by privacy preserving data release methods. 

\section{Privacy Preserving Data Sharing}
Just as modern data collection and sharing grew and evolved over the past few decades more modern versions of privacy protections did as well. Initially, data scrubbing techniques such as k-anonymity and l-diversity scrubbed data tables to ensure that individual records in the released data could not be linked to identities of persons. \textit{K}-anonymity, for example, guaranteed that for each row in the dataset at least $k-1$ rows shared the exact same identifiers in an attempt to ensure that any one row can not be uniquely linked to an identity.  As data analysis methodology became more advanced privacy attacks grew in sophistication along with them. Data scrubbing techniques were shown to be ineffective when presented against additional information; even information coming from other anonymized sources. For example, under \textit{k}-anonymity,  two tables could individually be 5-anonymized. However, when combined they could contain enough unique information to identify individuals that are in both tables. Likewise Dinur and Nissim \cite{dinur_nissim_2003} showed that too many aggregate statistics (even noisy ones) about a population could be used to reconstruct all the attributes of the underlying population. It is this theory that the US Census Bureau leveraged in its demonstration of the reconstruction attack in 2019. 
\par 

As a result, differential privacy has emerged as a privacy protection standard that a variety of algorithms can satisfy. Unlike table scrubbing techniques, differentially private mechanisms can only release  aggregated statistics protected by noise infusion. In order to avoid the reconstruction attacks of Dinur and Nissim \cite{dinur_nissim_2003} differentially private mechanisms bound the amount of data they release (and their accuracy) through a privacy loss parameter $\epsilon$, commonly referred to as the privacy loss budget. Higher values of $\epsilon$ allow for more queries to be answered and at higher quality levels resulting in less privacy. Meanwhile, low values of $\epsilon$ only allow a small number of queries to be answered with high noise infusions resulting in higher privacy. Unlike the table scrubbing techniques, differential privacy also had the additional property of \textit{privacy composition}: i.e., multiple releases of the same data using different privacy   also satisfied differential privacy albeit with a weaker privacy guarantee. This allows a data curator to track privacy loss as multiple privacy algorithms are run on the data.

\begin{figure}[t]
\centering
\includegraphics[width= 0.6 \textwidth]{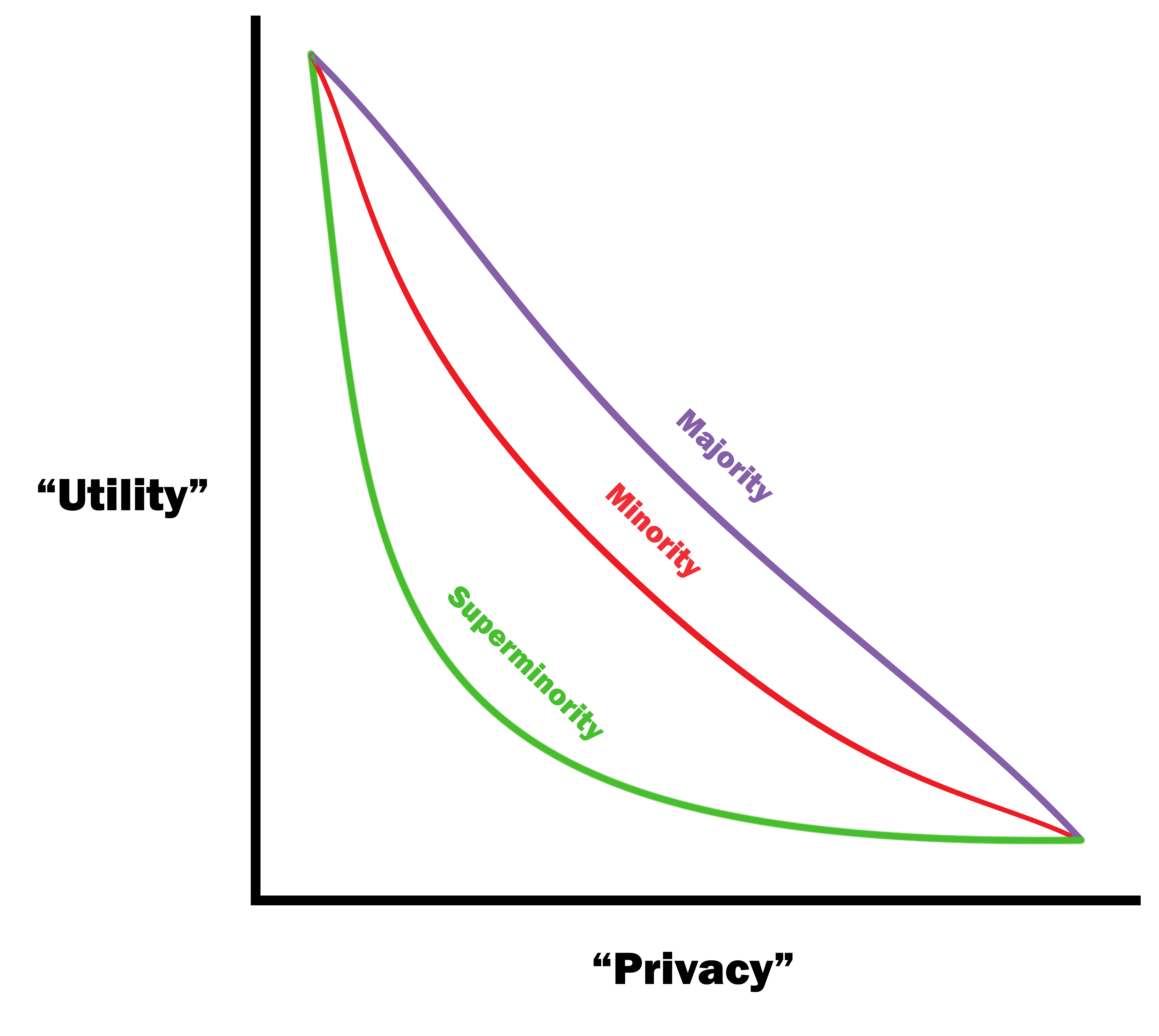}
\caption{Privacy Utility Tradeoff of Different Sized Populations}
\label{fig:example}
\end{figure}
\section{Privacy \& Equity}
Privacy however doesn't come for free. Each of the privacy mechanisms come at a cost to the quality of data and what can be represented in it. For \textit{k}-anonymity, the higher the privacy protections the less granular a data release can be. Even differential privacy pays this cost. Under differential privacy, as the privacy guarantees get stronger, (privacy budget is reduced) either fewer queries can be released about the data, and/or the data released is noisier. This presents a problem in data analysis: as privacy algorithms hide the information of individuals they also dilute and redact information relevant to small groups. These same privacy protections that are intended for individual privacy may end up erasing properties of small minority groups. Consider the extreme case where an individual is part of a group small enough that  identification in that group would be enough information to identify the individual. By necessity, any privacy preserving mechanism would need to erase that association, be it by scrubbing it away or by injecting noise larger than the signal from that group.\par 
Though rare, these circumstances do occur. Consider the language of Holikachuk, a native Alaskan language. This language is nearly extinct and contains only 5 known speakers \cite{AlaskanNatives}. Any attempt at protecting the privacy of these individuals must inherently erase their status as Holikachuk speakers. This phenomenon is not unique to superminorities, it applies to most minority groups that are underrepresented in data. As privacy protections get stronger, the data of larger minorities must inherently be diluted and erased in order to protect their privacy. 
In the literature, we usually see this as a tradeoff between privacy and \textit{fairness} (a term that encompasses several notions of fair treatment of individuals and groups). However, we think it's more appropriate to think of it specifically as a conflict between privacy and equal treatment across groups, which we will call \textit{equity}.
This leaves us with the following question, ``Will  the disproportionate affects of privacy protections on minority groups result in their unequal treatment from data driven decision making?''. 


%% file: Inequitites.tex
\section{Case Studies}

\textbf{Equity in Privacy Preserving Allocation:} Our Recent work \cite{PujolMKHMM20} investigates equity issues that might arise if policymaking (like funds allocations) were to use statistical data products released using differential privacy. In particular we looked at three allocation tasks whose mechanisms are well known -- Congressional Apportionment, Title 1 School fund allocations, and allocations of voting materials for minority languages. Each of these tasks can be expressed as simple functions (proportions and thresholds) computed from statistics released by agencies like the US Census Bureau. In each of the cases we found that if allocation tasks were executed on privacy protected statistics (as though they were the real statistics), there were significant disparities across different minority groups for each of the problems. 

For instance, the allocation of voting materials in native languages depends on whether or not a population group meets certain thresholds. When using differentially private statistics, while majority groups whose populations exceeded the threshold were rarely impacted, minority populations whose languages were often near the threshold (precisely the groups where accuracy is important) had a significantly higher rate of misclassification (not being allocated voting materials in native languages). 

In the case of Title 1 funding allocations the introduction of noise caused a biasing affect where small districts would receive drastically more funding with larger districts suffering as a result. For stringent settings of the privacy budget, this difference could be as much as 10 times the funding received by small districts and a 50\% loss experienced by large districts. However, the extent of inequity reduced for more realistic settings of the privacy loss budgets (e.g., like the ones that are used in deployments today). For both these problems the authors also found that the choice of the differential privacy algorithm used affected the level of inequity. In some cases, algorithms that introduced lower error overall resulted in higher inequity and vice versa. \par

\textbf{Equity in Privacy Preserving ML:} Bagdasaryan et al \cite{DPML_Disperate_Impact} investigate accuracy of privacy preserving machine learning on different subgroups of populations. They consider several machine learning tasks such as facial recognition, sentiment analysis, and species classification in nature images. For each of these tasks the authors created imbalanced training sets in order to enforce majority-minority groups. 
Note that algorithms for training ML models themselves have an equity/fairness problem. The authors in this work showed that training ML models under differential privacy further worsens the equity problem.

In the case of facial recognition, the authors showed that the difference in accuracy between individuals with light skin and those with dark skin is far worse in an ML model trained using a differentially private algorithm as opposed to a model trained without privacy constraints. The authors demonstrated that this was not a result of some faces being more difficult to classify than others but as a result of being in the minority. Across all the tasks examined the smaller the subgroup was within the training data the more differentially private training seemed to impact their classification accuracy. In sentiment analysis, tweets written in African American Vernacular English were misclassified more often than those written in standard American English. In facial recognition smaller age or gender groups were misclassified more often than their majority counterparts. Even in the case of nature image classification, a task devoid of the complexities of human speech or faces, there was a disparity in accuracy between the artificial minority and majority and this disparity only grew as the group size decreased or as the privacy protections increased. Even without other confounders simply being under-represented in the initial dataset could result in inequalities brought about by privacy protections. 

\textbf{Discussion:} Both the case studies discussed in this section showed that privacy mechanisms used to release statistics or ML models can introduce inequity if used as-is in downstream tasks. We will discuss in the next sections some ways to remedy the equity issues. We would like to note that we chose differential privacy as an exemplar for privacy methods, since it is mathematically rigorous, applies to a variety of data release modalities (statistics and ML), and potential biases can be theoretically analyzed (which may not true for legacy privacy techniques). That said, the privacy-equity tradeoff will arise no matter what privacy methodology is used. We would also like to note that privacy protection is just one of the reasons there might be bias in statistics \& ML models released from sensitive data. Other processing including imputation, record linkage, sampling, etc. also introduce biases/inequities. One nice aspect of differential privacy algorithms is that they add noise in a controlled manner thus making it easier to analyze and account for biases/inequities. 

Finally, another model of private data sharing that is getting a lot of press and traction in the industry is \textit{synthetic data}. These are \textit{fake} data generated to be in the same format as the original data but with the claim that they preserve statistical properties of the data while ensuring privacy of industries (often under differential privacy). Synthetic data methods work by learning a statistical or machine learnt model and then generate data as per that model. Thus, the privacy-accuracy and privacy-equity tradeoffs illustrated in prior work for privacy preserving statistics and ML also apply to synthetic data. More work is needed to systematically analyze this tradeoff for synthetic data given the hype around it. 

%% file: Resolution.tex
\section{Remedies}
From the examples above we can see that inequities can arise as a result of privacy protections in two different ways. In the case of allocation tasks using the privacy protected noisy data and treating it as true counts of populations can lead to various missallocations. In the case of machine learning tasks the naturally unbalanced groups with equal amounts of noise added to each of them results in minority groups which are subject to more noise and therefore more utility degradation. While this may seem as though privacy and equity are fundamentally at odds with one another this is not the case. Instead of assuming that the data given is true, allocation tasks can interpret their input as a noisy representation and account for it by either adding downstream repair steps or adjusting the allocation task entirely. Likewise there has been several new works, both theoretical and practical, which are leading to both fair and equitable ML tasks. Below we discuss strategies for accounting for the noise introduced by privacy mechanisms to mitigate the equity problem. 
\par 

\textbf{Noise Aware Allocation tasks: } Unlike other sources of noise, the distributions of the noise introduced by differential privacy are often known ahead of time. Allocations tasks can then leverage this information in order to adjust and avoid inequities. In the same paper where we demonstrate that differential privacy can cause inequities \cite{PujolMKHMM20} we suggest several downstream repair mechanisms. In the case of funding allocation problems, we suggest slightly increasing the overall allocation and distributing the additional budget in a way to counteract the biases introduced by the differentially private noise. This method leverages the known distribution of noise in order to ensure that no assignee (school districts, in their example of Title 1 funding) with high probability would receive less funds than in the non-private version of the same allocation task. Building on this case Fioretto et al. \cite{FairPrivFollowup} demonstrate that for a class of allocation functions (those where the entries of their hessian are constant functions) the equity loss due to privacy can be bounded. They show that the title one allocation is not in that class and offer an alternative solution where the allocation function is altered to have a known bounded fairness loss. \par 

\textbf{Equitable and Private ML: } Inequity in private ML is often due to the imbalance of the number of training examples from different sub-populations. One remedy for this is to simply collect more data from under-represented groups, therefore eliminating their minority status. Instead of eliminating the problem, rather, this results in minority groups falling under additional surveillance (or less privacy) to achieve the same level of accuracy as on majority populations. Ideally, we would like to design methods that can ensure ensure that ML models that are equitable both in terms of privacy as well as accuracy, and there is some recent research along that direction. Cummings et al. \cite{FairPrivCombatibility} theoretically proved that there does not exist a classifier which can satisfy ``pure"-differential privacy and satisfy several notions of equity while having higher utility than a constant classifier. Turns out this impossibility result does not hold under a weaker approximate version of differential privacy. 
Jagielski et al. \cite{DPFairLearning} have shown two algorithms which satisfy equalized odds (a metric for fair ML) while ensuring approximate differential privacy. Likewise, Tran et al.  \cite{DualApproach} develop a method of satisfying several group notions of equity and approximate differential privacy.

The aforementioned research suggests ways to mitigate the inequity introduced by privacy algorithms, but they just scratch the surface. We believe this is an important research area that is ripe for innovation and much more work is needed to understand privacy-equity trade-offs. 

\section{Conclusion}
In today's age of data collection on a massive scale, powerful tools to protect privacy are a necessity. These individual privacy protections inherently must necessarily hide properties of small groups of individuals. The current body of literature shows that this impacts decision processes and result in disparate impacts of utility/accuracy across different sub-populations. Privacy methods such as differential privacy provide us the mathematical tools to measure and account for the inequities introduced by the privacy methodology. As a remedy, we can adapt our decision processes to be aware of and account for the privacy protection methods and the inequities introduced by it. Theoretical exploration into the the fundamental trade-offs between privacy and equity can inform future methods for navigating the privacy-equity trade-off. 